\documentclass[fleqn,usenatbib]{mnras}

\UseRawInputEncoding
\usepackage{newtxtext,newtxmath}
\usepackage{gensymb}
\newcommand\target{{GX~339--4}}
\newcommand\nicer{{\it NICER}}

\newcommand\hxmt{{\it Insight}-HXMT}

\usepackage[T1]{fontenc}

\DeclareRobustCommand{\VAN}[3]{#2}
\let\VANthebibliography\thebibliography
\def\thebibliography{\DeclareRobustCommand{\VAN}[3]{##3}\VANthebibliography}

\usepackage{graphicx}	
\usepackage{amsmath}	
\usepackage{pdflscape}
\usepackage{verbatim}

\title[Fast transitions in GX 339--4]{Fast transitions of X-ray variability in the black hole transient GX 339--4: comparison with MAXI J1820+070 and MAXI J1348--630}

 \author[Zi-Xu Yang et al.]{
 Zi-Xu Yang,$^{1,2}$\thanks{E-mail: yangzx@ihep.ac.cn}
 Liang Zhang,$^{1}$\thanks{E-mail: zhangliang@ihep.ac.cn}
 S. N. Zhang,$^{1}$
 M. M\'{e}ndez,$^{3}$
 Federico Garc\'ia,$^{4}$
 Yue Huang,$^{1}$
 Qingcui Bu,$^{5}$
 \newauthor
 He-Xin Liu,$^{1,2}$
 Wei Yu,$^{1,2}$
 P. J. Wang,$^{1,2}$
 L. Tao,$^{1}$
 D. Altamirano,$^{6}$
 Jin-Lu Qu,$^{1}$
 S. Zhang,$^{1}$
 X. Ma,$^{1}$
 L. M. Song,$^{1}$
 \newauthor
 S. M. Jia,$^{1}$
 M. Y. Ge,$^{1}$
 Q. Z. Liu,$^{7}$
 J. Z. Yan,$^{7}$
 T. M. Li,$^{1,2}$
 X. Q. Ren,$^{1,2}$
 R. C. Ma,$^{1,2}$
 Yuexin Zhang,$^{3}$
 Y. C. Xu,$^{1,2}$
 \newauthor
 B. Y. Ma,$^{1,2}$
 Y. F. Du,$^{1,2}$
 Y. C. Fu,$^{1,2}$
 Y. X. Xiao,$^{1,2}$
 P. P. Li,$^{1,2}$
 P. Jin,$^{1,2}$
 S. J. Zhao$^{1,2}$
 and Q. C. Zhao$^{1,2}$
 \\
$^{1}$Key Laboratory for Particle Astrophysics, Institute of High Energy Physics, Chinese Academy of Sciences, 19B Yuquan Road, Beijing 100049, China\\
$^{2}$University of Chinese Academy of Sciences, Chinese Academy of Sciences, Beijing 100049, China\\
$^{3}$Kapteyn Astronomical Institute, University of Groningen, Postbus 800, 9700 AV Groningen, The Netherlands\\
$^{4}$Instituto Argentino de Radioastronomía (CCT La Plata, CONICET; CICPBA; UNLP), C.C.5, (1894) Villa Elisa, Buenos Aires, Argentina\\
$^{5}$Institut f\"ur Astronomie und Astrophysik, Kepler Center for Astro and Particle Physics, Eberhard Karls Universit\"at, Sand 1, D-72076 T\"ubingen, Germany」\\
$^{6}$Physics and Astronomy, University of Southampton, Southampton, Hampshire SO17 1BJ, UK \\
$^{7}$Purple Mountain Observatory, Chinese Academy of Sciences, Nanjing 210034, People's Republic of China
}
\date{Accepted XXX. Received YYY; in original form ZZZ}

\pubyear{2022}

\begin{document}
\label{firstpage}
\pagerange{\pageref{firstpage}--\pageref{lastpage}}
\maketitle

\begin{abstract}

Fast transitions between different types of power density spectra (PDS) happening over timescales of several tens of seconds are rare phenomena in black hole X-ray binaries. In this paper, we report a broadband spectral-timing analysis of the fast transitions observed in the 2021 outburst of \target\ using \nicer\ and \hxmt\ observations. We observe transitions between band-limited noise-dominated PDS and type-B quasi-periodic oscillations (QPOs), and their rapid appearance or disappearance. We also make a detailed comparison between the fast transitions in \target\ with those seen in MAXI J1820+070 and MAXI J1348--630. By comparing the spectra of the periods with and without type-B QPOs, we find that the spectral ratios above 10 keV are nearly constant or slightly decreasing, and the values are different between sources. Below 10 keV, the flux change of the Comptonization component is inversely proportional to the flux change of the thermal component, suggesting that the appearance of type-B QPOs is associated with a redistribution of the accretion power between the disc and the Comptonizing emission region. The spectral ratios between the periods with type-B QPO and those with broadband noise are significantly different from that with type-B QPO and without type-B QPO, where the ratios (type-B QPO/broadband noise) show a maximum at around 4 keV and then decrease gradually towards high energies. Finally, we discuss the possible change of the geometry of the inner accretion flow and/or jet during the transitions.

\end{abstract}




\section{Introduction}

Low-mass black hole X-ray binaries (BHXBs) consist of a stellar-mass black hole accreting matter from its companion star. These systems spend most of the time in a quiescent state, and occasionally go into bright outbursts due to the hydrogen-ionization instability in the outer part of the accretion disc \citep{Lasota2001}. Black hole transients in outburst usually evolve through different spectral states with changing luminosity \citep[see][for a review]{2006csxs.book..157M}. At the beginning of the outburst, a source is usually detected in a hard state (HS), where the energy spectrum is dominated by a Comptonization component with a power-law photon index of $\sim$1.4--1.7 \citep{1997ApJ...479..926M,2004PThPS.155...99Z}. As the accretion rate increases, the thermal component gradually strengthens and the source evolves into the intermediate state (IMS). We can further divide the IMS into a hard-intermediate state (HIMS) and a soft-intermediate state (SIMS), according to the spectral and timing properties \citep{2005Ap&SS.300..107H}. When the thermal component dominates the spectrum, the source enters into a soft state (SS), where the inner radius of the disc is thought to reach the innermost stable circular orbit (ISCO, e.g., \citealt{2007A&ARv..15....1D}). At the end of the outburst, the source fades quickly and moves back to the HS via the IMS in reverse order \citep{2016ASSL..440...61B}.

For BHXBs, not only the spectral shape but also the timing properties change significantly during an outburst. Fourier analysis provides a simple and intuitive way to inspect the fast timing variability. Narrow and strong peaks with frequencies ranging from a few mHz to $\sim$30 Hz, called low-frequency quasi-periodic oscillations (LFQPOs), are usually detected in the power density spectra (PDS) of these sources. Based on the properties of the PDS, LFQPOs can be divided into three types, i.e., type-A, B and C \citep{2005ApJ...629..403C}. Among the three types of QPOs, type-C QPOs are the most common type mostly appearing in the HS and HIMS. These QPOs are usually accompanied by strong flat-top noise variability. Although the physical origin of the type-C QPOs is still under debate, some evidence shows that they likely originate from the Lense-thirring precession of the inner hot flow or jet \citep{2009MNRAS.397L.101I,2010MNRAS.405.2447I,2011MNRAS.415.2323I,2016MNRAS.460.3284K,2018ApJ...866..122H,2021ApJ...919...92B,2021NatAs...5...94M,2022NatAs...6..577M}. However, this model faces considerable challenges both theoretically and observationally \citep{2021ApJ...906..106M,2022MNRAS.511..255N}. \citet{2022MNRAS.512.2686Z} fitted the time-dependent Comptonization model to the type-C QPOs in MAXI J1535--571. Type-B QPOs have been detected in a number of sources; the appearance of these QPOs defines the start of the SIMS \citep{2016ASSL..440...61B}. The type-B QPOs were found to occur close in time to the launch of discrete jet ejecta \citep{ 2004MNRAS.355.1105F,2009MNRAS.396.1370F,2020ApJ...891L..29H,2021MNRAS.504..444C}. Using phase-resolved spectroscopy, \citet{2016MNRAS.460.2796S} found that the spectral parameters change as a function of QPO phase, leading to a geometric interpretation for type-B QPOs. \citet{2020A&A...640L..16K} proposed a precessing jet model, which could reproduce the periodic variation of $\Gamma$ with QPO phase.
\citet{2021MNRAS.501.3173G} and \citet{2023MNRAS.519.1336P}, on the other hand, fitted the rms and lag spectra of the type-B QPO in MAXI J1348--630 and \target, respectively, with a time-dependent Comptonization model \citep{2020MNRAS.492.1399K,2022MNRAS.515.2099B}, in which the QPO properties are due to coupled oscillations of the corona and the disc. Type-A QPOs are the least common type of QPOs and have been rarely studied \citep{2019NewAR..8501524I}. No comprehensive model has been proposed to explain type-A QPOs.

Rapid transitions between different types of PDS happening over several tens of seconds have been observed in several BHXBs. Such transitions are usually observed in the IMS and always involve type-B QPOs. The study of these transitions can provide important evidence on what triggers the QPO phenomenon and on the physical origin of QPOs in general. 
Type-C QPO switching to/from type-B QPO is usually detected during the HIMS-SIMS state transitions, along with significant spectral variations \citep[e.g.,][]{2008MNRAS.383.1089S,2011MNRAS.418.2292M,2020ApJ...891L..29H}. Type-B/A QPO transitions have been found in GX 339--4 \citep{1991ApJ...383..784M,2003A&A...412..235N,2011MNRAS.418.2292M}, XTE J1550--564 \citep{2001ApJS..132..377H,2016ApJ...823...67S}, XTE J1859+226 \citep{2004A&A...426..587C,2013ApJ...775...28S}, XTE J1817--330 \citep{2012A&A...541A...6S} and H 1743--322 \citep{2005ApJ...623..383H}. 
%
%
In addition, LFQPOs are typically transient that can appear/disappear on short timescales of several seconds. 
%
%
%
Recently, \citet{2019ApJ...879...93X} reported a sudden turn-on of a transient type-C QPO accompanied by a rapid decrease in X-ray flux in Swift J1658.2--424.
\citet{2021MNRAS.505.3823Z} made a systematic analysis of the transient type-B QPO detected in MAXI J1348--630 using observations with the Neutron Star Interior Composition Explorer (\nicer). They found that the transient nature of the QPO is associated with a redistribution of the accretion power between the disc and the Comptonizing region. Meanwhile, \citet{2022ApJ...938..108L} fitted the broadband \hxmt\ spectra of periods with and without the type-B QPOs, and found a slight decrease in the height of the corona when the QPO disappears. 
%

%

\target\ is a BHXB with a mass function of $1.91\pm0.08 \ {\rm M}_{\sun}$ \citep{Heida2017}. 
%
%
It experienced frequent outbursts in the past two decades with a $\sim$2-yr cycle, making it one of the most studied stellar-mass black holes \citep{1986ApJ...308..635M,2004MNRAS.351..791Z}. \target\ entered a new outburst in 2021 \citep{2021ATel14336....1T,2021ATel14351....1P,2021ATel14352....1G,2021ATel14354....1S}, showing all four typical spectral states found in BHXBs. Both \nicer\ and \hxmt\ performed high-cadence observations of this outburst.
%
%
Using quasi-simultaneous data of \nicer\ and \hxmt, in this paper we report the study of the fast transitions between different types of PDS observed in the IMS of this source. The large effective areas and broad energy coverage of these observations with \nicer\ and \hxmt\ enable us to investigate in detail the changes of the spectral-timing properties during the fast transitions. Especially, we can constrain the disc, Comptonization and reflection components simultaneously by jointly fitting the \nicer\ and \hxmt\ spectra. We also compare the fast transitions in \target\ with those observed in the newly discovered BHXBs MAXI J1820+070 and MAXI J1348--630. 
%
%
In Section 2, we describe the observation and data reduction. The data analysis and main results are presented in Section 3. We discuss our main results in Section 4, and summarize our conclusions in Section 5.

\begin{table}
    \centering
        \caption{Log of the \nicer\ and \hxmt\  observations used in this paper. }
    \begin{tabular}{lccc}
        \hline\hline
        No.  & ObsID & Start time & Exposure (s)\\
        \hline
        \multicolumn{4}{c}{\nicer}\\
        \hline
        \#1 & 4133010105 & 2021 March 28 00:28:40 & 13605 \\
        \#2 & 4133010106 & 2021 March 29 02:38:28 & 10702 \\
        \#3 & 4133010107 & 2021 March 30 00:19:57 & 13793 \\
        \#4 & 4133010108 & 2021 March 31 01:07:00 & 10184 \\
        \hline
        \multicolumn{4}{c}{\hxmt}\\
        \hline
        \#1 & P030402403913 &  2021 March 29 12:22:02  & 1890 \\
        \#2 & P030402403914 &  2021 March 29 15:32:44 & 2490  \\
        \#3 & P030402403920 &  2021 March 30 13:13:50  & 2040 \\
        \#4 & P030402403921 &  2021 March 30 16:24:34  & 2580 \\
        \#5 & P030402403922 &  2021 March 30 19:35:18 & 2910  \\
        \#6 & P030402403923 &  2021 March 30 22:46:03  & 2520 \\
        \#7 & P030402403924 &  2021 March 31 01:56:47  & 2250 \\
        \#8 & P030402403925 &  2021 March 31 05:07:31  & 1140 \\
        \#9 & P030402403926 &  2021 March 31 08:18:16  & 1530 \\
        \#10& P030402403927 &  2021 March 31 11:29:00  & 1740 \\
        \#11& P030402403928 &  2021 March 31 14:39:45  & 2160 \\
        \#12& P030402403929 &  2021 March 31 17:50:29  & 2760 \\
        \#13& P030402403930 &  2021 March 31 21:01:13 &  2820 \\
         \hline
    \end{tabular}

    \label{log}
\end{table}


\begin{figure*}
    \centering
    \includegraphics[width=0.98\textwidth]{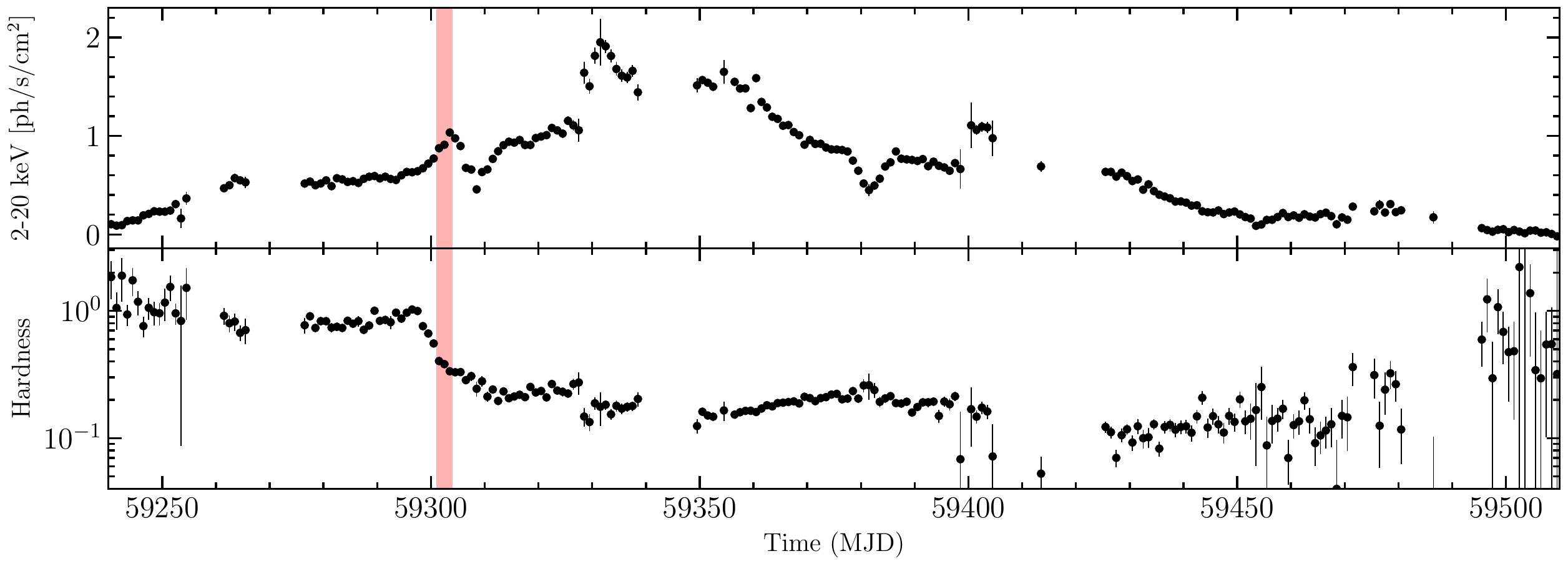}
    \caption{The evolution of 2--20 keV count rates and hardness ratio (4--10 keV/2--4 keV) of GX 339--4 in its 2021 outburst captured by MAXI/GSC. The red region marks the time interval we studied in this paper.}
    \label{MAXI_lc}
\end{figure*}

\begin{figure*}
    \centering
    \includegraphics[width=0.98\textwidth]{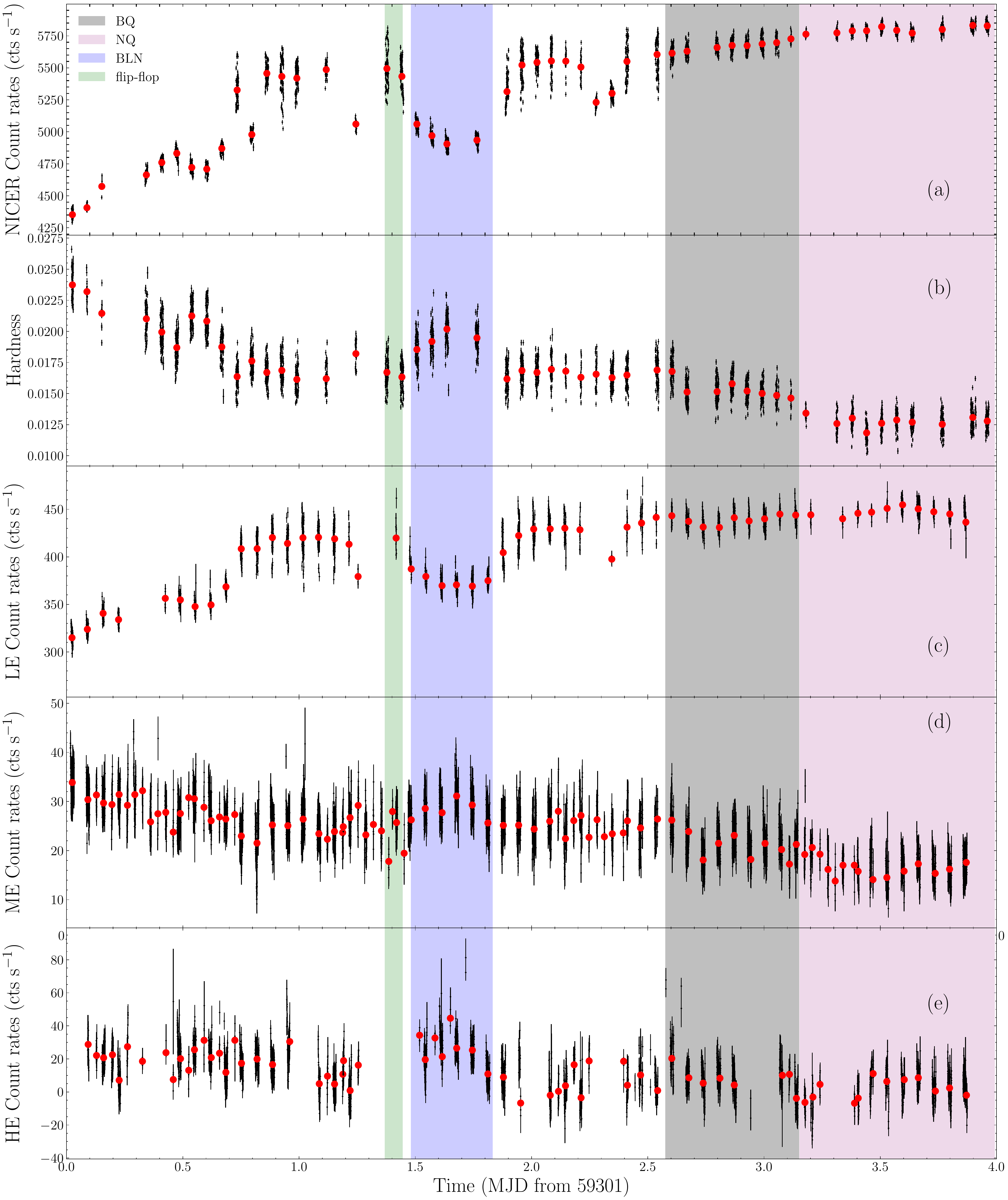}
    \caption{Evolution of the \nicer\ 0.5--10 keV light curve, hardness ratio (6.0--10.0 keV/2.0--4.0 keV), together with the \hxmt\ LE (1.0--10 keV), ME (8.0--35.0 keV), and HE (35.0--200.0 keV) light curves of \target\ for the period we study. The different shaded regions mark the time intervals we used to create the PDS and energy spectra for different types of variability (see the text for details).}
    \label{lc}
\end{figure*}

\begin{figure*}
    \centering
    \includegraphics[width=0.98\textwidth]{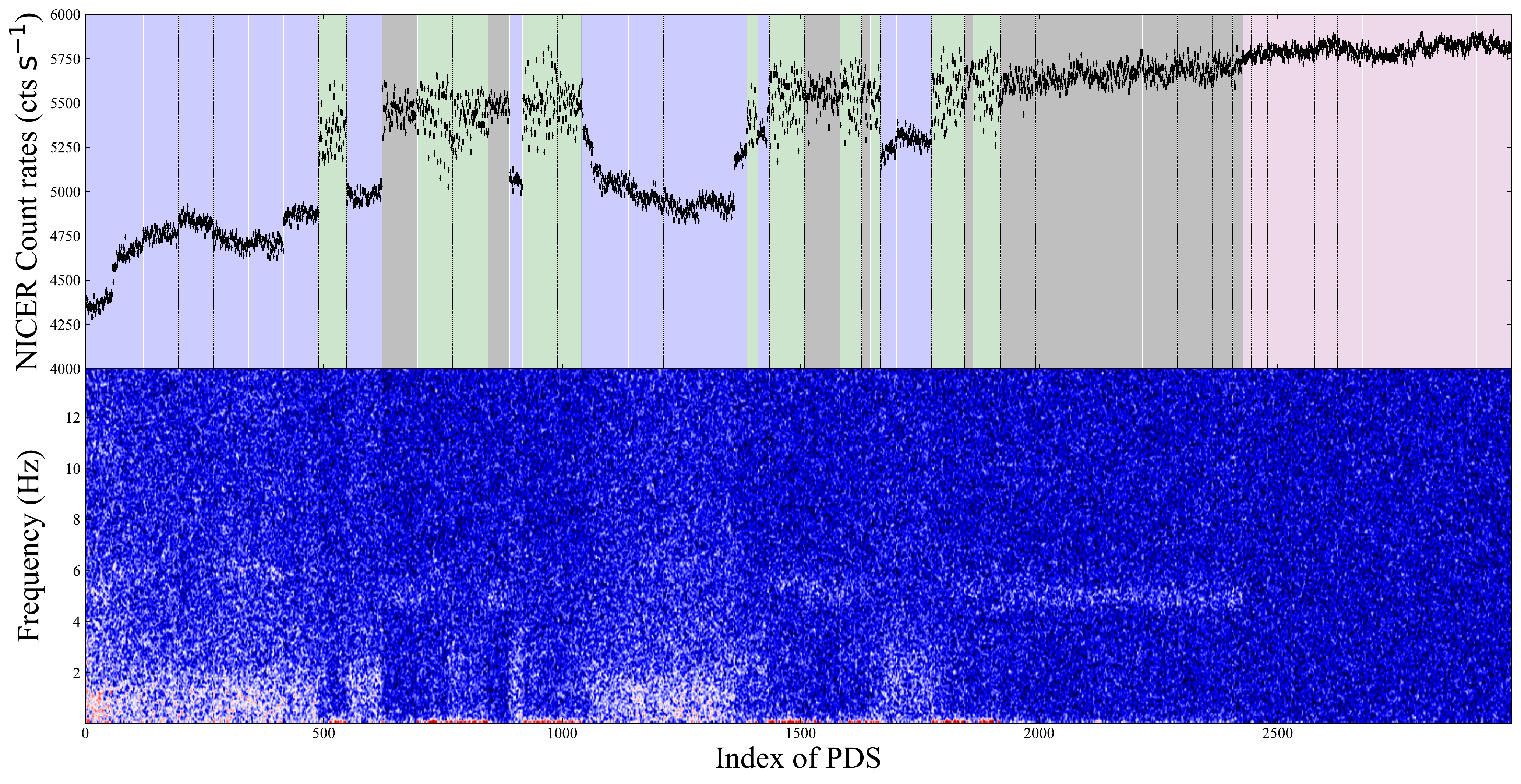}
    \caption{\nicer\ 0.5--10 keV light curve with 16-s binning (upper panel) and corresponding dynamical power density spectrum (lower panel) of \target\ for the period we study. All time gaps are removed and marked with dotted lines in the upper panel. Four different types of variability are identified based on the light curve and timing properties. The blue, gray and green shaded regions mark the periods with a relatively strong broadband noise component, a type-B QPO, and the flip-flop phenomenon in the PDS, respectively. The magenta shaded region represents the periods when the PDS are dominated by Poisson noise (see the text for details).}
    \label{lc_dps}
\end{figure*}


\begin{figure*}
    \centering
    \includegraphics[width=0.98\textwidth]{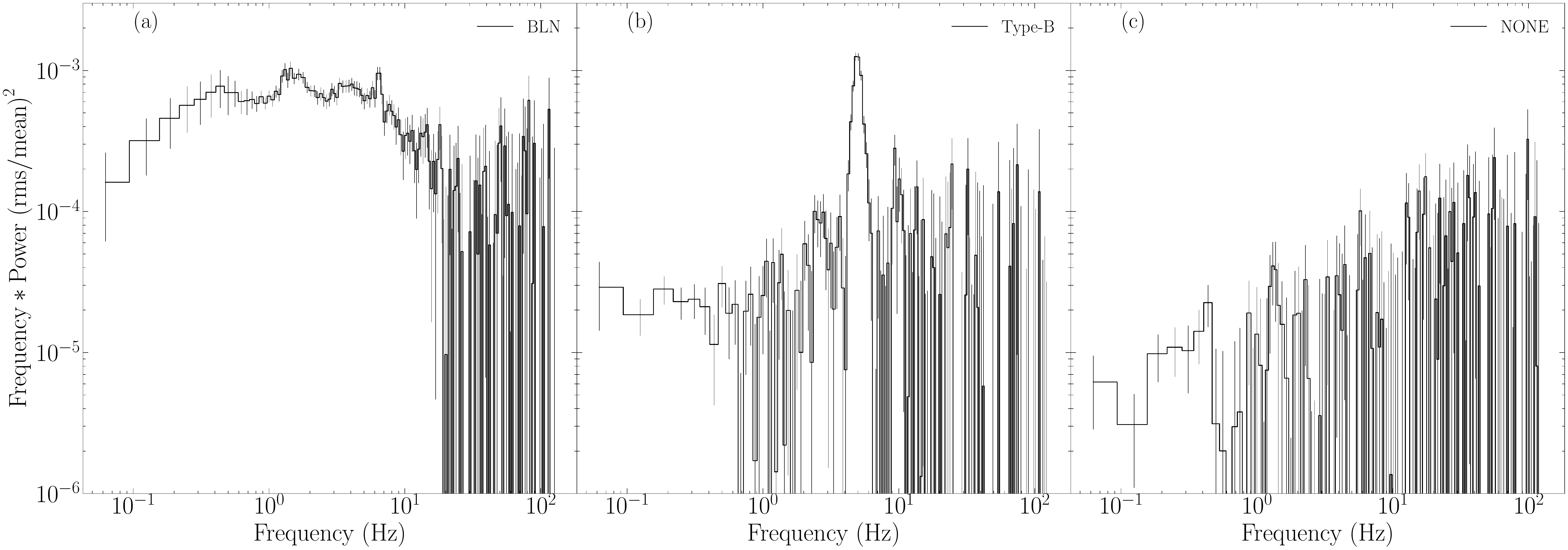}
    \caption{Representative power density spectra of \target\ using the \nicer\ 0.5--10 keV band. \textit{Panel a}: PDS with a strong band-limited noise component, extracted from the time interval MJD 59302.50--59302.77. \textit{Panel b}: PDS with a typical type-B QPO, extracted from the time interval MJD 59303.60--59304.12. \textit{Panel c}: PDS dominated by Poisson noise without any peaks, extracted from the time interval MJD 59304.18--59304.97.}
    \label{PDS_BLN_typeB_none}
\end{figure*}


\begin{figure*}
    \centering
    \includegraphics[width=0.98\textwidth]{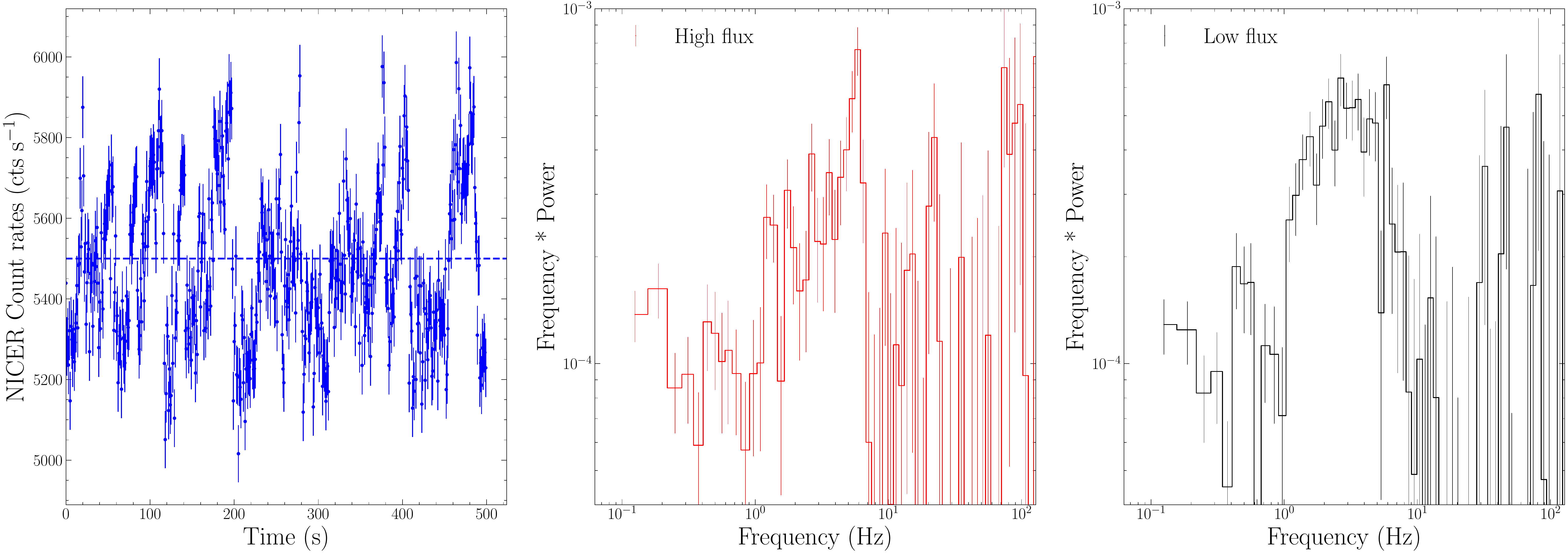}
    \caption{Left panel: a representative light curve of the flip-flop variability in \target\ with 1-s binning. The data are extracted from the time interval MJD 59302.37--59302.44. We divided the light curve into a high-flux and a low-flux interval based on the blue dashed line. The time origin is MJD 59302.3710. Middle and right panels: the average power spectra for the high-flux (red) and low-flux interval (black) of the flip-flop period in \target. The data used to produce the power spectra are shown in Fig.~\ref{lc} with green.}
    \label{flip_flop_high_low}
\end{figure*}

\begin{figure*}
    \centering
    \includegraphics[width=0.98\textwidth]{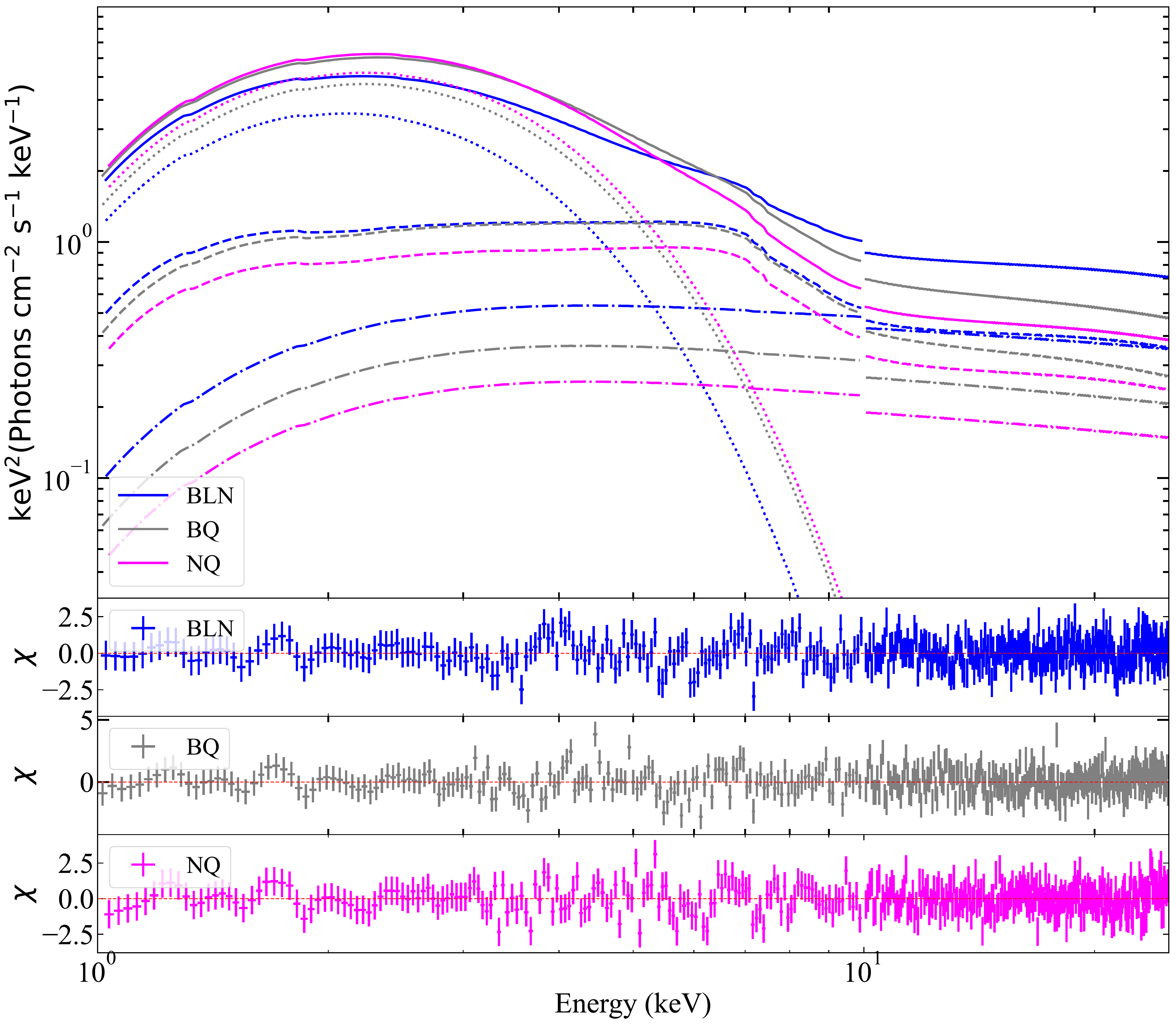}
    \caption{Best-fitting spectral model and residuals for the spectra of \target\ in the period with broadband noise (blue), type-B QPO (gray) and without any QPO (magenta), fitted with model 1: \textsc{constant*Tbabs*(diskbb+nthcomp+nthratio*relxillCp)}.}
    \label{spec_BLN_BQ_NQ_relxillCp}
\end{figure*}

\begin{figure*}
    \centering
    \includegraphics[width=0.98\textwidth]{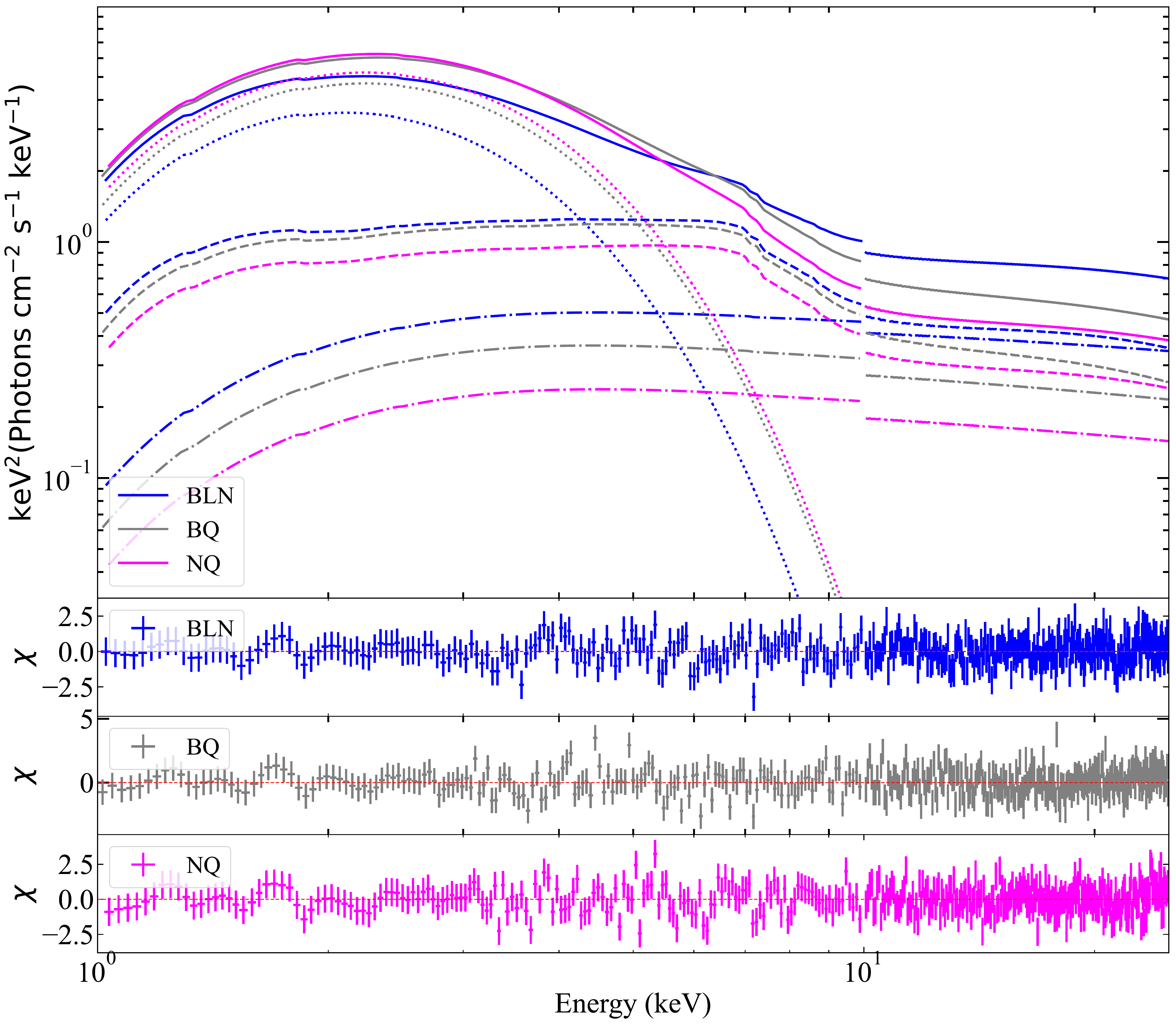}
    \caption{Best-fitting spectral model and residuals for the spectra of \target\ in the period with broadband noise (blue), type-B QPO (gray) and without any QPO (magenta), fitted with model 2: \textsc{constant*Tbabs*(diskbb+nthcomp+nthratio*relxilllpCp)}.}
    \label{spec_BLN_BQ_NQ_relxilllpCp}
\end{figure*}

\begin{table*}
    \centering
    \caption{Best-fitting spectral parameters for the spectra of \target\ in the period with broadband noise (BLN), type-B QPO (BQ), and without any QPO (NQ), fitted with model 1: \textsc{constant*Tbabs*(diskbb+nthcomp+nthratio*relxillCp)}. The spectra are combined from multiple consecutive orbits showing a similar PDS shape to improve the signal-to-noise ratio. The time intervals used to create the spectra are shown in Fig.~\ref{lc} with different colors (blue for BLN, gray for BQ and magenta for NQ).
    %
    }
    \begin{tabular}{llccc}
    \hline\hline
    Component & Parameters & BLN & BQ & NQ\\
    \hline
    Constant & {} & $0.89\pm{0.01}$ & $0.85\pm{0.01}$ & $0.85\pm{0.01}$ \\
    \hline
    Tbabs & $N_{\rm H} (\times10^{22}\rm cm^{-2})$ & \multicolumn{3}{c}{$0.46\pm{0.01}$}\\
    \hline
    Diskbb & $kT_{\rm in} ({\rm keV})$ & $0.71\pm{0.01}$ & $0.77\pm{0.01}$ & $0.78\pm{0.01}$\\
    {}     & norm                      & $2551^{+108}_{-53}$ & $2342^{+82}_{-62}$ & $2556^{+63}_{-51}$\\
    \hline
    Nthcomp & $\Gamma$ & $2.16\pm{0.01}$ & $2.21\pm{0.01}$ & $2.20\pm{0.01}$ \\
    {}      & $kT_{\rm e} (\rm keV)$ & \multicolumn{3}{c}{100 (fix)}  \\
    {}      & norm         & $0.29^{+0.04}_{-0.02}$ & $0.19\pm{0.06}$ & $0.13\pm{0.04}$ \\
    \hline
    RelxillCp & $i~(\degree)$            & \multicolumn{3}{c}{$34.14\pm{1.00}$}\\
    {}        & $a$                      & \multicolumn{3}{c}{0.998 (fix)}\\
    {}        & $R_{\rm in}({\rm ISCO})$           & $-1.55\pm{0.06}$ & $-1.63\pm{0.06}$ & $-1.71^{+0.13}_{-0.08}$\\
    {}        & $R_{\rm out}(R_{\rm g})$          & \multicolumn{3}{c}{400 (fix)}\\
    {}        & z                      & \multicolumn{3}{c}{0 (fix)}\\
    {}        & Index          & $3.50\pm{0.09}$ & $3.58\pm{0.10}$ & $3.45\pm{0.09}$\\
    {}        & log ($\xi$)             & $2.99\pm{0.03}$ & $2.99\pm{0.04}$ & $3.00\pm{0.06}$\\
    {}        & log$N$ (${\rm cm}^{-3}$)                   & $20.00^{+P^{\dag}}_{-0.05}$ & $20.00^{+P^{\dag}}_{-0.08}$ & $19.73^{+0.17}_{-0.22}$\\
    {}        & $A_{\rm Fe}$           & \multicolumn{3}{c}{$3.34^{+0.87}_{-0.50}$}\\
    {}        & $kT_{\rm e} ({\rm keV})$ & \multicolumn{3}{c}{100 (fix)}\\
    {}        & norm                   & $0.013\pm{0.001}$  & $0.012\pm{0.001}$  & $0.010\pm{0.001}$\\
    \hline
        Flux              &                        &                         &                         &   \\
    $F_{\rm diskbb}$   &   ($10^{-8} \ {\rm erg} \ {\rm cm}^{-2} \ {\rm s}^{-1}$)                     & $0.87\pm{0.01}$                        &        $1.17\pm{0.01}$                 &   $1.31\pm{0.01}$ \\
    $F_{\rm nthcomp}$  &    ($10^{-8} \ {\rm erg} \ {\rm cm}^{-2} \ {\rm s}^{-1}$)                    &             $0.24\pm{0.01}$            &    $0.16\pm{0.01}$                     &   $0.11\pm{0.01}$  \\
    $F_{\rm relxillCp}$ & ($10^{-8} \ {\rm erg} \ {\rm cm}^{-2} \ {\rm s}^{-1}$)                    &         $0.53\pm{0.01}$                &    $0.49\pm{0.01}$                     &  $0.39\pm{0.01}$    \\
    \hline
    $R_{\rm f}$   &   &  $2.21\pm{0.10}$  & $3.06\pm{0.20}$   & $3.54\pm{0.33}$  \\
    \hline
    {}        & $\chi^2/d.o.f.$        & \multicolumn{3}{c}{1026/1202} \\
    \hline\hline
    \multicolumn{5}{|l|}{$^\dag$: The letter $(P)$ indicates that the error of the parameter pegged at the upper boundary.} \\
    
    \end{tabular}
    \label{table_spec_sep_nthratio_relxillCp_link12}
\end{table*}

\begin{table*}
    \centering
    \caption{Best-fitting spectral parameters for the spectra of \target\ in the period with broadband noise (BLN), type-B QPO (BQ), and without any QPO (NQ), fitted with  model 2: \textsc{constant*Tbabs*(diskbb+nthcomp+nthratio*relxilllpCp)}. The spectra are combined from multiple consecutive orbits showing a similar PDS shape to improve the signal-to-noise ratio. The time intervals used to create the spectra are shown in Fig.~\ref{lc} with different colors (blue for BLN, gray for BQ and magenta for NQ). 
    %
    }
    \begin{tabular}{llccc}
    \hline\hline
    Component & Parameters & BLN & BQ & NQ\\
    \hline
    Constant & {} & $0.89\pm{0.01}$ & $0.85\pm{0.01}$ & $0.85\pm{0.01}$ \\
    \hline
    Tbabs & $N_{\rm H}(\times10^{22}\rm cm^{-2})$ & \multicolumn{3}{c}{$0.46\pm{0.01}$}\\
    \hline
    Diskbb & $kT_{\rm in} ({\rm keV})$ & $0.71\pm{0.01}$ & $0.77\pm{0.01}$ & $0.78\pm{0.01}$\\
    {}     & norm                      & $2592^{+71}_{-52}$ & $2351^{+43}_{-34}$ & $2586^{+37}_{-30}$\\
    \hline
    Nthcomp & $\Gamma$ & $2.14\pm{0.02}$ & $2.19\pm{0.01}$ & $2.18\pm{0.02}$ \\
    {}      & $kT_{\rm e} ({\rm keV})$ & \multicolumn{3}{c}{100 (fix)}  \\
    {}      & norm         & $0.27\pm{0.04}$ & $0.18\pm{0.03}$ & $0.12\pm{0.03}$ \\
    \hline
    RelxilllpCp & $i~(\degree)$            & \multicolumn{3}{c}{$29.90^{+0.94}_{-1.38}$}\\
    {}        & $a$                      & \multicolumn{3}{c}{0.998 (fix)}\\
    {}        & $R_{\rm in}({\rm ISCO})$           & $-1.71\pm{0.03}$ & $-1.84\pm{0.04}$ & $-1.87\pm{0.07}$\\
    {}        & $R_{\rm out}(R_{\rm g})$          & \multicolumn{3}{c}{400 (fix)}\\
    {}        & $h (R_{\rm g})$          & $1.49^{+0.32}_{-0.21}$ & $1.49^{+0.11}_{-0.20}$ & $1.96^{+0.32}_{-0.38}$\\
    {}        & z                      & \multicolumn{3}{c}{0 (fix)}\\
    {}        & log($\xi$)             & $2.99\pm{0.04}$ & $2.98\pm{+0.03}$ & $3.01^{+0.06}_{-0.04}$\\
    {}        & log$N$ (${\rm cm}^{-3}$)                   & $20.00^{+P^{\dag}}_{-0.08}$ & $20.00^{+P^{\dag}}_{-0.08}$ & $19.94^{+P^{\dag}}_{-0.26}$\\
    {}        & $A_{\rm Fe}$           & \multicolumn{3}{c}{$3.51^{+0.59}_{-0.41}$}\\
    {}        & $kT_{\rm e} ({\rm keV})$ & \multicolumn{3}{c}{100 (fix)}\\
    {}        & norm                   & $2.91^{+0.77}_{-1.79}$  & $2.66^{+1.56}_{-1.24}$  & $0.30^{+0.63}_{-0.13}$\\
    \hline
    Flux              &                        &                         &                         &   \\
    $F_{\rm diskbb}$   &   ($10^{-8} \ {\rm erg} \ {\rm cm}^{-2} \ {\rm s}^{-1}$)                     & $0.89\pm{0.01}$                        &        $1.20\pm{0.01}$                 &   $1.31\pm{0.01}$ \\
    $F_{\rm nthcomp}$  &    ($10^{-8} \ {\rm erg} \ {\rm cm}^{-2} \ {\rm s}^{-1}$)                    &      $0.23\pm{0.01}$                   &           $0.16\pm{0.01}$              &  $0.10\pm{0.01}$   \\
    $F_{\rm relxilllpCp}$ & ($10^{-8} \ {\rm erg} \ {\rm cm}^{-2} \ {\rm s}^{-1}$)                    &         $0.54\pm{0.01}$                &        $0.50\pm{0.01}$                 &   $0.40\pm{0.01}$   \\
    \hline
    $R_{\rm f}$  &     &  $2.34\pm{0.11}$  &  $3.12\pm{0.20}$  & $4.00\pm{0.41}$  \\
    \hline
    {}        & $\chi^2/d.o.f.$        & \multicolumn{3}{c}{993/1202} \\
    \hline\hline
    \multicolumn{5}{|l|}{$^\dag$: The letter $(P)$ indicates that the error of the parameter pegged at the upper boundary.} \\
    \end{tabular}

    \label{table_spec_sep_nthratio_relxilllpCp}
\end{table*}

\section{OBSERVATIONS AND DATA Reduction}

The MAXI/GSC light curve and the evolution of the hardness ratio are shown in Fig.\ref{MAXI_lc}. The hardness ratio evolves from  a high value of $\sim$1 to a low value of $\sim$0.2 and back to a high value at the end of the outburst. This suggests that the source experienced a full outburst with an initial hard-to-soft transition and a soft-to-hard transition. We refer to \citet{2023MNRAS.519.1336P} for the detailed evolution track observed with \nicer. \nicer\ observed that the flux of \target\  exhibited a dramatic increase from $\sim$2500 counts s$^{-1}$ on 2021 Mar 26 (MJD 59299) to $\sim$4600 counts s$^{-1}$ on 2021 Mar 28 (MJD 59301; \citealt{2021ATel14490....1W}). Multi-wavelength observations indicated that \target\ was already in the HIMS, and was evolving rapidly towards the SIMS \citep{2021ATel14490....1W,2021ATel14493....1B,2021ATel14507....1C,2021ATel14504....1L}. Both \nicer\ and \hxmt\ conducted very intense observation campaigns during this epoch that covered a complete state transitions between the HIMS and the SIMS. During the state transition, we observed fast alternations between different types of PDS. In this paper, we focus on the observations during this transition period; the logs of the observations for \nicer\ and \hxmt\ we used in this paper are listed in Table~\ref{log}. 

%

\nicer\ is a soft X-ray telescope onboard the International Space Station (ISS), launched on 2017 June 3. It provides high throughput in the 0.2--12 keV energy band with an absolute timing precision of $\sim$100 ns, making it an ideal instrument to study fast X-ray variability. \nicer\ consists of 56 co-aligned concentrator X-ray optics, each paired with a single-pixel silicon drift detector. Presently, 52 detectors are active with a peak effective area of $\sim$1900 $\rm cm^{-2}$ at 1.5 keV. In our analysis, we exclude data from detectors \#14 and \#34 as they occasionally show episodes of increased instrumental noise.
We processed the \nicer\ observations using the \nicer\ software tools NICERDAS version 8.0 in HEASOFT version 6.29.  
We run the pipeline {\tt nicerl2} with the standard filtering criteria.

\hxmt\ is China's first X-ray astronomy satellite, launched on 2017 June 15 \citep{2020SCPMA..6349502Z}.
It carries three slat-collimated instruments: the High Energy X-ray telescope (HE, poshwich NaI/CsI, 20--250 keV, \citealt{2020SCPMA..6349503L}), the Medium Energy X-ray telescope (ME, Si pin detector, 5--30 keV, \citealt{2020SCPMA..6349504C}), and the Low Energy X-ray telescope (LE, SCD detector, 1--15 keV, \citealt{2020SCPMA..6349505C}).
The \hxmt\ observations are extracted from all three instruments using the \hxmt\ Data Analysis software (HXMTDAS) v2.05\footnote{The data analysis software is available from \url{http://hxmten.ihep.ac.cn/software.jhtml}.}, and filtered with the following criteria:  
(1) pointing offset angle less than $0.04\degree$;  
(2) Earth elevation angle larger than $10\degree$;  
(3) geomagnetic cutoff rigidity larger than 8 GeV;  
(4) at least 300 s before and after the South Atlantic Anomaly passage.
Each instrument of \hxmt\ carries large, small and blind field-of-view (FoV) detectors in order to facilitate background analysis. The small FoV detectors (LE: 1\degree.6 $\times$ 6\degree; ME: 1\degree $\times$ 4\degree; HE: 1\degree.1 $\times$ 5\degree.7) have a lower probability of source contamination. Therefore, to avoid possible contamination from the bright Earth and nearby sources, we only use data from the small FoV detectors \citep{2018ApJ...864L..30C}.

\section{Analysis and Results}

\subsection{Fast transitions between different types of PDS}

In Fig.~\ref{lc} we plot the evolution of the \nicer\ 0.5--10 keV count rate, hardness ratio (6.0--10.0 keV/2.0--4.0 keV), along with the \hxmt/LE 1.0--10.0 keV, ME 8.0--35.0 keV, and HE 35.0--200.0 keV light curves of the period we study. The data of \hxmt/HE are dominated by background, so that we did not use them for our further analysis. %
From this figure, it is apparent that the source exhibited dramatic flux changes, mainly jumping between a low- and a high-flux state, accompanied by significant changes in spectral hardness.

For our timing analysis, we create dynamical and average PDS in the \nicer\ 0.5--10.0 keV energy band with 16-s data segments and 1/256-s time resolution. We normalised the PDS to ${\rm rms}^2/{\rm Hz}$ \citep{1990A&A...230..103B,1991ApJ...383..784M}, and subtracted the contribution due to Poisson noise. In this work we only show the timing results of the \nicer\ data, since the \nicer\ count rate is an order of magnitude higher than that of \hxmt\ and, therefore, the features, i.e., broadband noise or QPOs, are more significant in the PDS of \nicer.

In Fig.~\ref{lc_dps}, we show the \nicer\ 0.5--10 keV light curve with 16-s binning and the corresponding dynamical PDS of the same period as in Fig.~\ref{lc}. All time gaps are removed in this plot and marked with dotted lines. The $x$-axis represents the index of each 16-s PDS. The same plot of each individual observation is shown in Fig.~\ref{lc_dps_appendix}.
From the dynamical PDS, it is apparent that the shape of the PDS varies significantly between different orbits, coinciding with changes in flux and spectral hardness (see also Fig.~\ref{lc}). Based on their different timing characteristics, we can divide the PDS of this period into three types.
%
%
In Fig.~\ref{PDS_BLN_typeB_none}, we show the representative PDS of each type. In the upper panel of Fig.~\ref{lc_dps}, we have marked the time of the PDS shown in Panels (a), (b) and (c) with blue, gray and magenta, respectively.

The PDS as the one shown in Panel (a) of Fig.~\ref{PDS_BLN_typeB_none} usually appear at a low-flux state when the count rate is below $\sim$5200 count s$^{-1}$, corresponding to a higher hardness ratio. These PDS are dominated by a relatively strong broadband noise component with a 0.1--128 Hz fractional rms amplitude above 5\% (hereafter we call `BLN' to this type of PDS). Sometimes weak type-C QPOs (not always visible) are detected at around 5--7 Hz. This type of PDS is consistent with that typically observed at the end of the HIMS \citep{2011MNRAS.418.1746S,2020ApJ...891L..29H}. 
The PDS as the one shown in Panel (b) are detected at a high-flux state, corresponding to a slightly softer spectral hardness. Their overall rms amplitude value, $\sim$2\% in the 0.1--128 Hz, is significantly lower than that of the PDS shown in Panel (a). A sharp narrow 5-Hz QPO (quality factor\footnote{$Q=\nu_0/{\rm FWHM}$, where $\nu_0$ is the centroid frequency of the QPO and FWHM is the full width at half maximum.} $Q\sim7$) is seen with a fractional rms amplitude of $\sim$1.7\%. The centroid frequency of the QPO remains more or less constant with time. Based on these characteristics, we can identify this as a type-B QPO (hereafter we call `BQ' to this type of PDS). The type-B QPOs are usually observed when a source enters the SIMS \citep{2016ASSL..440...61B}. 
Still at a high-flux level, when the type-B QPO disappears, we observe the PDS as the one shown in Panel (c). In this case, the energy spectrum is softer than in the periods shown in Panels (a) and (b). The PDS is dominated by Poisson noise without any significant peak. The 0.1--128 Hz fractional rms is below 1\%. This is the typical PDS observed in the SS of BHXBs (hereafter we call `NQ' to this type of PDS). 

%

It is important to mention that we also observed a peculiar kind of light curve in some orbits of this period (green in Fig.~\ref{lc_dps}):  the source undergoes rapid transitions between a low- and a high-flux state on timescales of several tens of seconds. In the left panel of Fig.~\ref{flip_flop_high_low}, we show a representative light curve of this type. This type of variability, called flip-flops, has been observed in previous outbursts of \target\ \citep{1991ApJ...383..784M} and in some other sources, e.g.,  Swift J1658.2-4242 \citep{2020A&A...641A.101B},  GS 1124-683 \citep{1997ApJ...489..272T} and XTE J1550-564 \citep{2001ApJS..132..377H,2016ApJ...823...67S} .
We then split the light curve into a high-flux (above the blue dashed line) and a low-flux interval (below the blue dashed line). The PDS of the two intervals are shown in the middle and right panels of Fig.~\ref{flip_flop_high_low}.
We find that the PDS of the high-flux interval is similar to that shown in Panel (b) of Fig.~\ref{PDS_BLN_typeB_none}. A significant 5-Hz QPO is seen with weak broadband noise. The PDS of the low-flux interval is dominated by a low-frequency noise component. 

\subsection{Energy spectra corresponding to different types of PDS}

We extract representative energy spectra corresponding to the different types of PDS shown in Fig.~\ref{PDS_BLN_typeB_none}. For our spectral analysis, we use \nicer\ for the low-energy spectral coverage and \hxmt/ME for the high-energy spectral coverage. We do not use the \hxmt/HE data due to the high background contribution. We create the \nicer\ background-subtracted spectra with the Background Estimator Tool `nibackgen3C50'\footnote{\url{https://heasarc.gsfc.nasa.gov/docs/nicer/tools/nicer_bkg_est_tools.html}}. We binned the spectra in accordance with the optimal binning algorithm proposed by \citet{Kaastra2016} with a minimum of 50 counts per spectral bin. We then added a systematic error of 1\% percent below 3.0 keV using GRPPHA. We produced the \hxmt/ME spectra using the HXMT pipeline. The energy bands adopted for spectral analysis are 1.0--10.0 keV for \nicer\ (we ignore the data below 1.0 keV due to the considerable instrumental residuals) and 10--25 keV for \hxmt/ME. We fit the three representative spectra jointly using XSPEC version 12.12.10 \citep{1996ASPC..101...17A}. 
%

%


For our spectral fitting, we first use a simple continuum model consisting of an absorbed disc component and a Comptonization component, \textsc{constant*Tbabs*(diskbb+nthcomp)} in XSPEC. The \textsc{constant} reflects the relative calibration between \nicer\ and \hxmt/ME, and is fixed to 1 for \nicer\ and left free to vary for \hxmt/ME. We find a strong and broad Fe K emission line centered at 6--7 keV in the residuals, accompanied by a weak reflection hump above 10 keV.
%
%
We, therefore, include a full reflection model in our fitting \citep{2014ApJ...782...76G,2014MNRAS.444L.100D,2016A&A...590A..76D,2022MNRAS.514.3965D}.  We have tried the relativistic reflection models within an empirical broken power-law emissivity ({\sc relxillCp}) and that with the emissivity inferred from the lamp-post geometry ({\sc relxilllpCp}). 
It is worth noting that both {\sc relxillCp} and {\sc relxilllpCp} calculate the relativistic disc reflection spectra using seed photon spectra of {\sc nthcomp} fixed at 0.01 keV. This is compatible with a cutoff power-law, and very different from the typical illuminating spectra ({\sc diskbb} temperature $\sim$ 0.5 keV) observed in BHXBs, thus introducing an artificial soft-excess in the modeling. To improve this issue, we use the multiplicative model {\sc nthratio}\footnote{\url{https://github.com/garciafederico/nthratio}}, which aims to correct for this soft excess by introducing a first-order phenomenological correction, renormalising the reflected spectrum by the ratio between the {\sc nthComp} models at $kT_{\rm bb}$ and the fixed value of 0.01~keV. The model depends on three parameters, i.e., $\Gamma$, $kT_{\rm e}$ and $kT_{\rm bb}$, which we link to $\Gamma$, $kT_{\rm e}$, $kT_{\rm bb}$ from the fitted {\sc nthcomp} and {\sc diskbb} components. This effectively does not add extra parameters to the model.
%
%
Therefore, our final models to describe the spectra are model 1: \textsc{constant*Tbabs*(diskbb+nthcomp+nthratio*relxillCp)}, and model 2: \textsc{constant*Tbabs*(diskbb+nthcomp+nthratio* \
relxilllpCp)}. 
%
%
%

In both models, we linked the hydrogen column density ($N_{\rm H}$), inclination and the iron abundance together between the spectra of the three epochs as they are not likely to change in such a short period. Since \target\ is found to harbor a near maximally spinning black hole \citep{2015ApJ...806..262L,2016ApJ...821L...6P}, we fix the value of the spin at $a_{*}=0.998$ to check the possible change of the inner disk radius.
%
%
%
%
Since the high-energy cutoff of the hard component is outside the energy range of the instrument, we fix the electron temperature ($kT_{\rm e}$) at 100 keV. We separate the disc reflection component and the illuminating Comptonization component by fixing the reflection fraction ($R_{\rm f}$) at $-1$ and we calculate $R_{\rm f}$ from the flux of the reflection and the Comptonization component we measured.
In model 1, we assume a single power-law emissivity profile. Using a broken power-law emissivity profile results in overfitting of the spectra. 
In model 2, the effect of returning radiation on the shape of the X-ray reflection spectrum is considered \citep{2022MNRAS.514.3965D}.
The other parameters in both models are allowed to vary between spectra. Both models give an acceptable spectral fit with $\chi^2/d.o.f. = 1026/1202$ for model 1 and $\chi^2/d.o.f. = 993/1202$ for model 2. The best-fitting spectral models and residuals are shown in Fig.~\ref{spec_BLN_BQ_NQ_relxillCp} and Fig.~\ref{spec_BLN_BQ_NQ_relxilllpCp}, and the best-fitting spectral parameters are listed in Table~\ref{table_spec_sep_nthratio_relxillCp_link12} and Table~\ref{table_spec_sep_nthratio_relxilllpCp}.

From Tables~\ref{table_spec_sep_nthratio_relxillCp_link12} and \ref{table_spec_sep_nthratio_relxilllpCp}, it is apparent that the evolution of the main spectral parameters is very similar between the two models. The inner disc radius, $R_{\rm in}$, measured from the reflection spectrum is consistent within errors between the spectra of the three epochs. However, the normalisation of the disc component varies moderately. The change of the {\sc diskbb} normalisation could be due to the change of the spectral hardening factor. By fitting the \hxmt\ broadband spectra of the 2021 outburst of \target,  \citet{2022arXiv221109543L} found that the inner disc radius remains stable during the HIMS-SIMS state transition. If $R_{\rm in}$ does not change, the change of the {\sc diskbb} normalisation could be due to the change of the spectral hardening factor \citep{1998PASJ...50..667K}.  In this case, we find that the spectral hardening factor in the BLN, BQ and NQ epochs is $1.25\pm{0.02}$, $1.32\pm{0.02}$, $1.32^{+0.03}_{-0.05}$ for model 1, and $1.32\pm{0.01}$, $1.40\pm{0.02}$, $1.38\pm{0.03}$ for model 2 by assuming $i = 30\degree, D = 8 \ {\rm kpc}, M_{\rm BH} = 10\ M_{\odot}$. Therefore, minor changes in the spectral hardening factor could account for the changes of the {\sc diskbb} normalisation. The change of the spectral hardening factor could be due to the change of the disc density \citep{2022ApJ...927..210Z} or the illumination of the hard X-rays from the corona onto the disc \citep{2022ApJ...932...66R}.

From BLN to BQ, the inner disc temperature increases with a slight spectral softening.
The reflection fraction increases gradually from BLN to BQ to NQ. The height of the corona measured from model 2 does not change too much between the spectra of the three epochs. However, we note that their values ($h\sim 1.5-2.0 R_{\rm g}$) are very small, which would indicate that the illuminating source is very close to the black hole.

\begin{figure}
    \centering
    \includegraphics[width=0.49\textwidth]{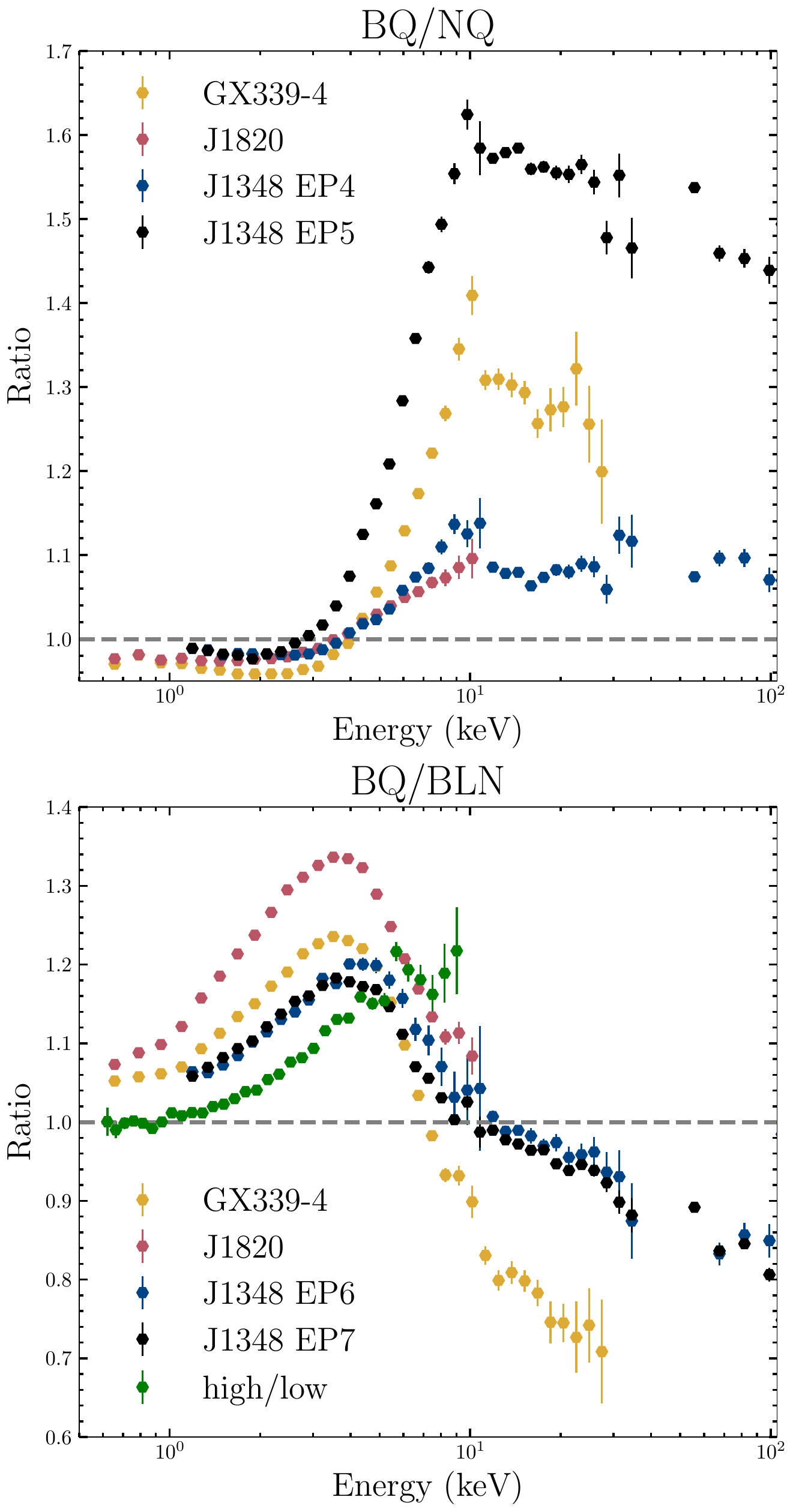}
    \caption{For \target, MAXI J1820+070 and MAXI J1348--630, upper panel: ratio between the spectrum with and without the type-B QPO (BQ/NQ), and lower panel: ratio between the spectrum with the type-B QPO and that with the broadband noise (BQ/BLN). The green points mark the spectral ratio between the high-flux level and the low-flux level of the flip-flop variability. The data of MAXI J1348--630 are adopted from \citet{2022ApJ...938..108L}. The data points are re-binned for clarity.}
    \label{spec_ratio}
\end{figure}

\begin{figure}
    \centering
    \includegraphics[width=0.49\textwidth]{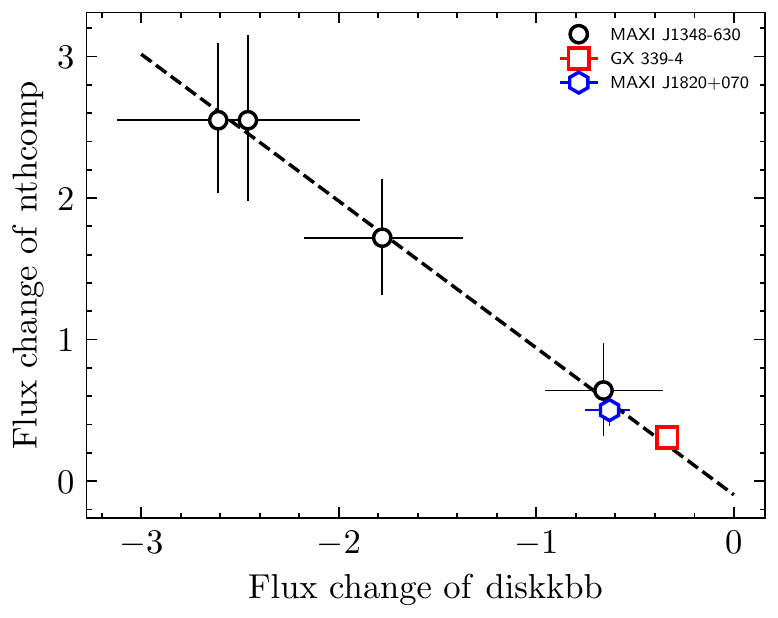}
    \caption{The change of the Comptonization component flux as a function of the change of the disk component flux from non-QPO to type-B QPO for \target, MAXI J1820+070 and MAXI J1348--630. The fluxes are calculated in the 0.6--10.0 keV and are in units of $10^{-8} \ {\rm erg}\ {\rm cm}^{-2} \ {\rm s}^{-1}$. The dashed line represents the best-fitting line $\Delta F_{\rm nthcomp}= -1.03 \Delta F_{\rm diskbb} - 0.09$.}
    \label{delta_flux}
\end{figure}

\begin{figure}
    \centering
    \includegraphics[width=0.49\textwidth]{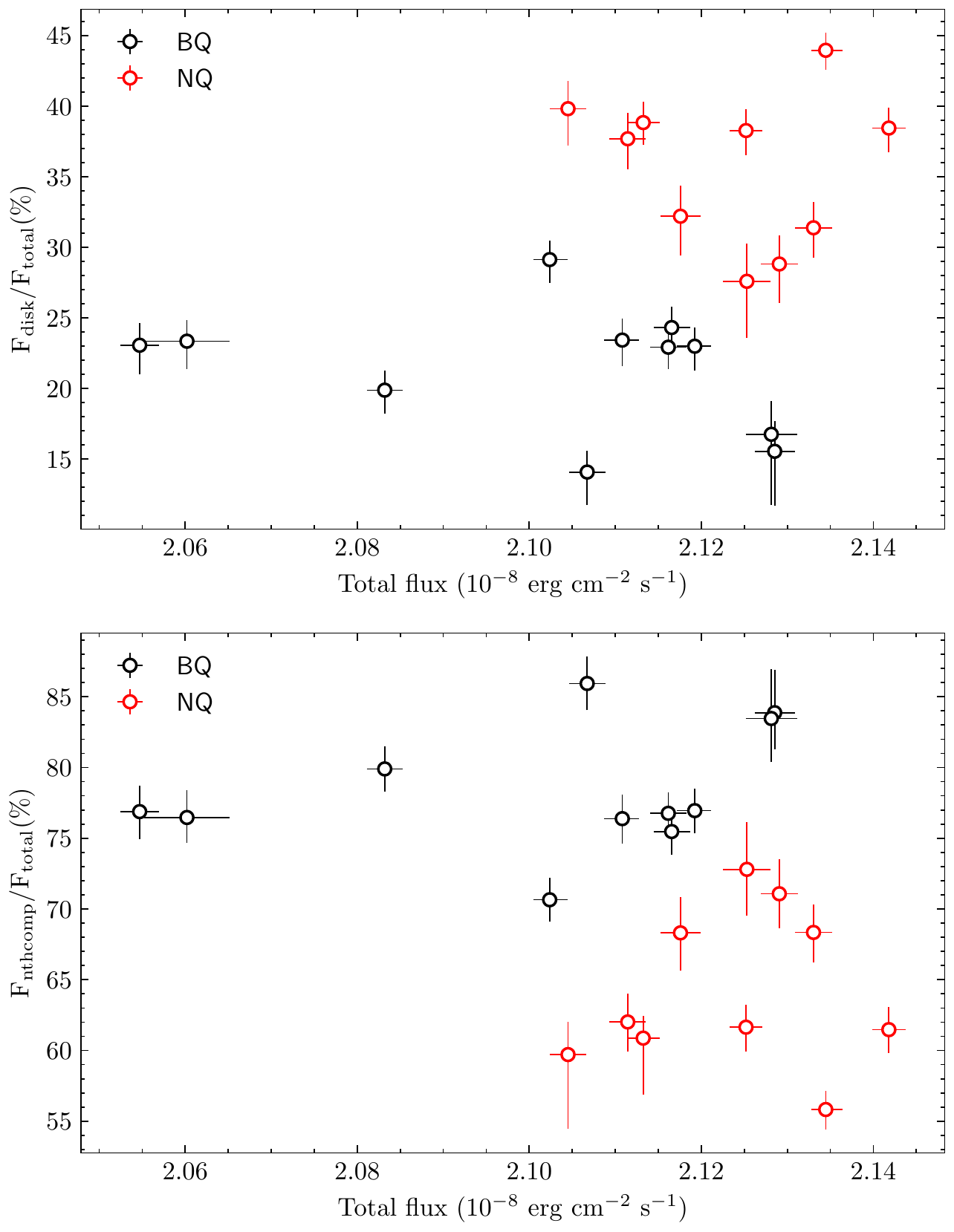}
    \caption{Flux fraction of the {\sc diskbb} component ($F_{\rm disk}/F_{\rm total}$, upper panel) and the {\sc nthcomp} component ($F_{\rm nthcomp}/F_{\rm total}$, lower panel) as a function of the total flux of \target\ for each orbit with (black) or without (red) the type-B QPO.
    \label{flux_ratio}}
\end{figure}

\subsection{Comparison with MAXI J1820+070 and MAXI J1348--630}

In this section, we compare the transitions we observed in \target\ with those seen in the two bright BHXBs MAXI J1820+070 and MAXI J1348--630.

%

A rapid HIMS-to-SIMS state transition was observed in MAXI J1820+070 by \nicer\ (ObsID: 1200120197), along with fast changes in the X-ray variability \citep{2020ApJ...891L..29H}. A type-C QPO seen at the beginning of the transition became gradually weaker and was replaced by a type-B QPO at some point. About 2--2.5 hours after this switch, a strong radio flare was observed that corresponded to the launch of jet ejecta \citep{2020ApJ...891L..29H}. The type-B QPO was only present for $\sim$1.5 hours and then disappeared. For our analysis, we processed this observation with the standard filtering criteria using NICERDAS version 8.0. We then extracted background-subtracted spectra for the period with type-C QPO (BLN), type-B QPO (BQ), and without any QPOs (NQ), respectively. We fitted the \nicer\ spectra with the model \textsc{Tbabs*(diskbb+nthcomp+gauss)} in the 1.5--10.0 keV band. The data below 1.5 keV are not used due to the large instrumental residuals seen in the spectra. Unfortunately, \hxmt\ did not conduct observations during this transition.

For MAXI J1348--630, fast transitions between different kinds of PDS were detected by \nicer\ \citep{2021MNRAS.505.3823Z} and \hxmt\  \citep{2022ApJ...938..108L}. 
Using \hxmt\ data, \citet{2022ApJ...938..108L} detected multiple cases of BQ-NQ and BLN-BQ transitions in this source and presented a detailed spectral analysis in a broad energy band of 1--100 keV. \citet{2021MNRAS.505.3823Z} reported the detection of a fast appearance/disappearance of a type-B QPO (BQ-NQ transition) in four \nicer\ observations and performed the first systematic spectral-timing analysis of this transition. 
The results of MAXI J1348--630 shown in this paper are adopted from \citet{2022ApJ...938..108L} and \citet{2021MNRAS.505.3823Z}.


%

In the upper panel of Fig.~\ref{spec_ratio}, we show the ratio between the spectra with and  without the type-B QPO (BQ/NQ). The data of MAXI J1348--630 are adopted from the upper panel of fig.6 of \citet{2022ApJ...938..108L}.
The overall trend of the spectral ratio is quite similar between the three black hole systems. The main spectral differences appear at energies above 3 keV, where the ratio increases towards higher energies and turns to flat or slightly decreases with energy above 10 keV. The values of the ratio above 10 keV are different between sources and between different cases of the same source.
The spectral ratios below 3 keV are all lower than unity, suggesting a weaker disc component when there is a type-B QPO than when there is no QPO.

The upper panel of Fig.~\ref{spec_ratio} shows that the high-energy emission is stronger when the type-B QPO appears. Then it is natural to consider that the type-B QPO is related to the appearance of an additional hard spectral component. To test this scenario, we perform a joint spectral fitting for the BQ and NQ spectra using a two-Comptonized component model, \textsc{constant*Tbabs*(diskbb+nthcomp1+nthcomp2+laor)} in XSPEC. In our fitting, the parameters of the first Comptonization component ({\sc nthcomp1}) including $\Gamma$, $kT_e$, $kT_{\rm bb}$, norm are linked between the BQ and NQ spectra. We assume that the type-B QPO is only produced in the second Comtonization component ({\sc nthcomp2}) and the normalization of this component is set to zero for the NQ spectrum. We obtain that the {\sc nthcomp2} component contributes $\sim$22\% of the flux to the BQ spectrum. If {\sc nthcomp1} does not contribute to the QPO rms amplitude, we measure that the intrinsic rms amplitude of {\sc nthcomp2} should be around 71\% to produce the observed QPO rms (16\% in the 20--30 keV band). We have also done a similar analysis for MAXI J1348--630 and the intrinsic rms of the additional hard component should be higher than 100\% to produce the 15\% rms of the type-B QPO. However, the rms of the variability in BHXBs is typically lower than 50\% \citep{2019NewAR..8501524I}. Therefore, it is not likely that only the increased hard component flux contributes to the QPO rms.

Based on a systematic analysis of the transient type-B QPO in MAXI J1348--630, \citet{2021MNRAS.505.3823Z} found that, in the \nicer\ energy band (0.6--10 keV), the change of the Comptonization component flux, $\Delta F_{\rm nthcomp}$, is anti-correlated with the change of the disc component flux, $\Delta F_{\rm diskbb}$, during the BQ-NQ transition (see their fig. 8). This suggests that the appearance of the type-B QPO is related to a redistribution of the accretion power between the disc and the Comptonization component. In Fig.~\ref{delta_flux}, it is apparent that the three sources, \target, MAXI J1820+070 and MAXI J1348--630, follow the same correlation. We fit this correlation using a linear function and find the best-fitting line is $\Delta F_{\rm nthcomp}= -1.03 \Delta F_{\rm diskbb} - 0.09$, consistent with the result of \citet{2021MNRAS.505.3823Z} obtained using the data of MAXI J1348--630 only.

By comparing the spectra of the orbits with and without type-B QPO, \citet{2021MNRAS.505.3823Z} found a clear threshold in the fraction of the Comptonization component flux to the total flux ($F_{\rm nthcomp}/F_{\rm total}$, 0.6--10 keV band), which is related to the appearance of type-B QPO: the QPO only appears when $F_{\rm nthcomp}/F_{\rm total} \gtrsim 40\%$.
Following the method described in \citet{2021MNRAS.505.3823Z}, we make a similar analysis for \target\ using \nicer\ data. In Fig.~\ref{flux_ratio} we show $F_{\rm disk}/F_{\rm total}$ (top panel) and $F_{\rm nthcomp}/F_{\rm total}$ (bottom panel) against the total flux. Here each point corresponds to one orbit from the BQ/NQ epoch in Fig.~\ref{lc_dps}. We also find a clear threshold in $F_{\rm nthcomp}/F_{\rm total}$ or $F_{\rm diskbb}/F_{\rm total}$ in this plot. The QPO is present when $F_{\rm nthcomp}/F_{\rm total} \gtrsim 70\%$.  
%

In the lower panel of Fig.~\ref{spec_ratio}, we show the ratio between the spectra with type-B QPO and those with the broadband noise (BQ/BLN). The data of MAXI J1348--630 are adopted from the lower panel of fig.6 of \citet{2022ApJ...938..108L}.
The spectral ratios of BQ/BLN are quite different from those of BQ/NQ. Below $\sim$10 keV, the ratios are greater than unity, and first increase and then decrease with energy, showing a maximum at around 4 keV. Above $\sim$10 keV, the ratios are lower than unity and decrease towards higher energies up to 100 keV with a different slope. 
We also extract a spectrum for the high-flux level and the low-flux level of the flip-flop variability, respectively. The spectral ratio (high-flux/low-flux) is shown in the lower panel of Fig.~\ref{spec_ratio} with green points. We find that below $\sim$4 keV, the spectral ratio increases monotonously with energy, similar to that of BQ/BLN. However, above $\sim$4 keV, the evolution of the spectral ratio is quite different. The spectral ratio between the high-flux and low-flux level of the flip-flops remains more or less constant without showing a decreasing trend.
In addition, we note that no anti-correlation between the change of the thermal and Comptonization component flux is seen during the BLN/BQ transition. This can be inferred from the evolution of the spectral ratio with energy easily as the spectral ratio is always greater than unity below 10 keV.

\section{discussion}

Fast transitions in X-ray variability have been reported in several sources during the past three decades. Thanks to the large effective area of \nicer\ and \hxmt, we can study, for the first time, in detail the spectral and variability changes during the fast transitions in a broad energy band. 
%
%
In this paper, we have investigated the changes of the spectral-timing properties during the fast transitions observed in \target, and made a detailed comparison of the transitions between different black hole systems. 
%
%
We find that the spectral ratio between intervals with a type-B QPO (BQ) and without a QPO (NQ) is nearly constant or slightly decreasing above 10 keV, and the value of the constant is different between sources.
Below 10 keV, a change in the relative contribution of the disc and Comptonization emission was observed during the BQ and NQ transitions. The type-B QPOs were only detected when the fraction of the Comptonization component is above a threshold. The value of the threshold is found to be different between sources.  
In addition, the trend of the spectral ratio is quite different between the two different kinds of transition (BQ/NQ and BQ/BLN). 
Below we discuss our main findings.

\subsection{The transient nature and physical origin of the type-B QPO}

Type-B QPOs are typically transient features that suddenly appear/disappear on short timescales \citep{2003A&A...412..235N,2013ApJ...775...28S}. In this paper, we have made a detailed spectral-timing study of the transient type-B QPOs detected in three BHXBs. By comparing the spectra of the periods with and without the QPO, we find that the overall trend of the spectral ratio is very similar between sources. This similarity suggests that the transient QPOs in BHXBs might share the same physical origin.

As shown in the top panel of Fig.~\ref{spec_ratio}, the spectral ratio between BQ and NQ for each case is nearly constant or slight decreasing above 10 keV, and the value is significantly higher than unity. This means that the high-energy Comptonization emission is stronger when type-B QPOs appear. \citet{2022ApJ...938..108L} found that the value above 10 keV is different between different cases in a source. In contrast, the total flux below 10 keV does not change too much. If the flux could be an indicator of mass accretion rate, our result implies that the appearance of type-B QPOs or not is not determined by the change of the global mass accretion rate only, given that the BQ/NQ transition can occur at different flux levels in an outburst \citep{2021MNRAS.505.3823Z,2022ApJ...938..108L}.

At energies below 10 keV, the spectral change between BQ and NQ is more complex. We find that, in all cases, the Comptonization component flux increases whereas the disc component flux decreases from NQ to BQ. The flux change of the thermal component is inversely proportional to the flux change of the Comptonization component. This tight correlation supports the scenario proposed by \citet{2021MNRAS.505.3823Z} that the transient nature of the type-B QPO is associated with a redistribution of the accretion power between the disc and the Comptonization emission.

In MAXI J1348--630, \citet{2021MNRAS.505.3823Z} found that the type-B QPO only appears when the flux fraction of the Comptonization component is above a critical value. This threshold is also present here in \target. However, the critical value in \target\ is inconsistent with that obtained in MAXI J1348--630, suggesting that source properties determine this critical value.
%
%
The critical value may be related to the intrinsic parameters of the different BHXB systems (e.g., black hole mass, spin, inclination) or the different  configurations of the accretion flow. The inclination of GX 339--4 is $36 \degree \pm 4\degree$ (\citealt{2015ApJ...806..262L}) and the value is $36.5\degree \pm 1\degree$ for MAXI J1348--630 (\citealt{2022MNRAS.513.4869K}). Therefore, it is less likely that inclination plays an important role in determining the critical value. GX 339--4 contains an extremely fast-spinning black hole with $a_{*}>0.97$ \citep{2015ApJ...806..262L}, while MAXI J1348--630 has a moderately high spin ($a_{*} = 0.80 \pm 0.02$, \citealt{2022MNRAS.513.4869K}).
The mass of GX 339--4 is not well constrained ($2.3 \ {\rm M}_{\sun} < M_{\rm BH} < 9.5 \ {\rm M}_{\sun}$, \citealt{Heida2017}), whereas the mass of MAXI J1348--630 is $M_{\rm BH} = 8.7 \pm 0.3~{\rm M}_{\sun}$ \citep{2022MNRAS.513.4869K}. Given the current measurements of the intrinsic parameters of the two systems, it is difficult to distinguish whether black hole mass or spin value determines the threshold.

The type-B QPO observed in the 2021 outburst of \target\ was also studied by \citet{2023MNRAS.519.1336P} using \nicer\ and {\it AstroSat} data. They found that the rms and lag spectra of the type-B QPO in \target\ are similar to those in MAXI J1348--630 \citep{2021MNRAS.501.3173G}. In addition, by fitting the rms and lag spectra of the QPO with a time-dependent Comptonization model \citep{2022MNRAS.515.2099B}, these authors found that two physically-connected Comptonization components are needed to explain the radiative properties of the QPO: a small corona ($\sim25 R_{\rm g}$) with a high feedback fraction of photons returning from the corona to the disc and a large corona ($\sim1500 R_{\rm g}$) with almost no feedback \citep{2023MNRAS.519.1336P}. The small corona would be a compact region located close to the black hole, while the large corona would be a vertically extended region that may be the base of a relativistic jet or outflow. This supports our above finding that the increased hard component is not  the only contributor to the modulation of the QPO flux.


\subsection{Transitions between the PDS with type-B QPOs and strong broadband noise}

Fast transitions between the PDS with type-B QPOs and strong broadband noise (BQ/BLN) are usually observed during the HIMS-SIMS state transition \citep[e.g.,][]{2011MNRAS.418.2292M,2020ApJ...891L..29H,2020MNRAS.499..851Z}. Sometimes, a significant type-C QPO is detected when the broadband noise dominates. We found that the spectral ratio of this type of transition is significantly different from that of the BQ/NQ transition. As shown in the bottom panel of Fig.~\ref{spec_ratio}, the ratio shows a maximum at around 4 keV. Above 4 keV, the ratio decreases gradually with photon energy, changing from above unity to below unity at around 10 keV. We note that the decreasing trend is steeper below 10 keV.
 In addition, the different relation between thermal and Comptonization component flux during BLN/BQ transitions suggests that the physical process or the geometry leading to the two types of transitions is totally different.

%


In \target, we observed frequent BLN/BQ transitions during the period we study. Such phenomenon was also observed in the previous outbursts of \target\ and in some other BHXBs \citep{2016ASSL..440...61B}. In addition, BQ/BLN transitions are also detected during the flip-flop variability. This corresponds to frequent state transitions between the HIMS and the SIMS.
\citet{2014MNRAS.437.3994N} proposed a model to explain the state transitions in BHXBs considering the process of disc tearing. In this model, individual rings of gas in the disc break off and precess independently. The tearing of the disc and alignment with black hole spin can explain the frequent HIMS-SIMS transitions. However, this process happens on timescales of a few days \citep{2014MNRAS.437.3994N}. A previous study showed that the type-B QPO can appear/disappear within tens of seconds \citep{2003A&A...412..235N}, contrary to the expectation of the above model.

%
%

Based on a systematic study of the reverberation lags in 10 BHXBs, including the three sources we studied in this work, \citet{2022ApJ...930...18W} found that the lags of the broadband noise component in the PDS become longer during the HIMS-SIMS state transition, suggesting that the corona is the base of the jet that vertically expands and/or is ejected during the transition. \citet{2022MNRAS.512.2686Z} fitted simultaneously the time-averaged spectrum, the fractional rms and phase-lag spectra of the QPOs in MAXI J1535--571. The change of the geometry inferred from their results is that the corona expands vertically but contracts horizontally during the transition from the HIMS towards the SIMS. In this work, we find that the inner disk radius does not change too much from BLN to BQ. This also suggests that the spectral-timing evolution during the HIMS-SIMS state transition could be due to changes of the corona and/or geometry of the jet base. 
During the state transition from the HIMS to the SIMS, the jet properties change dramatically. The steady compact jet gradually switches off and relativistic discrete ejecta can be launched \citep[e.g.,][]{2009MNRAS.396.1370F,Russell2019,2021MNRAS.504..444C}. The evolution of the jet properties is also likely related to the change of the inner flow.


%

\subsection{Possible changes in corona/jet geometry during the transitions}

In the bright HS and the HIMS, type-C QPOs are usually present and accompanied by strong band-limited noise. The band-limited noise is supposed to originate from propagating fluctuations in mass accretion rate within a hot inner flow \citep{2016AN....337..385I}, and the type-C QPO can be produced by the global Lense-thirring precession of the same flow \citep{2009MNRAS.397L.101I}. The change of the projection area of the hot flow along the line-of-sight leads to the flux modulation of the QPO. However, the type-C QPO becomes weaker or non-detectable at the end of the HIMS. 
In the form of the advection-dominated accretion flow (ADAF, \citealt{1997ApJ...489..865E,2014ARA&A..52..529Y}), the gas cannot cool effectively and the accretion flow prefers a spherical accretion geometry (scale height $H/R\sim 1$). This symmetry becomes more dominant as the accretion rate increases in the outburst. Thus, the weakening of the type-C QPO could be due to the hot inner flow being very close to spherical so that the change of the projected area is minor, or the shrinking of the ADAF corona with increasing accretion rate.
During the transition from BLN to BQ, the band-limited noise component decreases significantly. From BLN to BQ, we find that the Comptonization component becomes softer and its flux decreases. Since the band-limited noise is usually related to the hot inner flow, we expect that the decrease of the band-limited noise is likely due to the shrinking of the hot flow region. 

As mentioned earlier, the type-B QPO could be produced in the combined extended corona and the base of the jet. 
%
%
%
\citet{2022ApJ...938..108L} proposed that the disappearance of the type-B QPO can be due to that the inner part of the flow becoming aligned with the black-hole spin axis due to the Bardeen-Petterson effect \citep{1975ApJ...195L..65B}. 
From our spectral fitting, we find that the reflection fraction increases slightly when the type-B QPO disappears. The reflection fraction in the lamppost geometry is related to the corona height and the bulk velocity \citep{2021NatCo..12.1025Y}. Since we observe a constant lamp post height during the transitions (Table~\ref{table_spec_sep_nthratio_relxilllpCp}), the change of the reflection fraction could be due to the change of the bulk velocity. 

\section{Conclusions}

We present a detailed spectral-timing analysis of the fast transitions of X-ray variability observed in the 2021 outburst of \target, and make a detailed comparison of the transitions between different black hole systems. The main results are:

(1) In the three black-hole X-ray binary systems \target, MAXI J1348--630 and MAXI J1820+070, the spectral ratios of BQ/BLN and BQ/NQ are very similar. The spectral ratio of BQ/NQ is nearly constant or slightly decreasing above 10 keV, but the value is different for each source. We also investigate the relative flux change of the thermal and Comptonization components during the appearance/disappearance of the type-B QPO and find a tight anti-correlation for the three sources with a slope $\sim$-1 below 10 keV.

(2) Using a broadband spectral analysis, we find that the inner radius of the accretion disc shows no evident changes during the transitions. However, the spectrum becomes softer when the type-B QPO appears. When the type-B QPO disappears, the reflection fraction increases.

We suggest that a slightly truncated disc with a hot flow region exists in the BLN epoch and the hot flow region converts into a jet that might be precessing when the type-B QPO appears. The disappearance of the type-B QPO can be explained by a stop of the precession of the jet. During the appearance/disappearance of the type-B QPO, the redistribution of the accretion power between the disc and the jet plays a key role.

\section*{Acknowledgements}

We thank the anonymous referee for constructive comments. This work is supported by NASA through the \nicer\ mission. This work has made use of the data from the \hxmt\ mission, a project funded by China National Space Administration (CNSA) and the Chinese Academy of Sciences (CAS), and data and/or software provided by the High Energy Astrophysics Science Archive Research Center (HEASARC), a service of the Astrophysics Science Division at NASA/GSFC. This work is supported by the National Key RD Program of China (2021YFA0718500) and the National Natural Science Foundation of China (NSFC) under grants 12203052, U1838201, U1838202, 11733009, 11673023, U1938102, U2038104, U2031205, 12233002, the CAS Pioneer Hundred Talent Program (grant No. Y8291130K2) and the Scientific and Technological innovation project of IHEP (grant No. Y7515570U1). This work was partially supported by International Partnership Program of Chinese Academy of Sciences (Grant No.113111KYSB20190020). M.M. acknowledges the research programme Athena with project number 184.034.002, which is (partly) financed by the Dutch Research Council (NWO). FG is a
CONICET researcher. FG acknowledges support by PIP 0113 (CONICET) PICT-2017-2865 (ANPCyT), and PIBAA 1275 (CONICET). YZ acknowledges support from the China Scholarship Council (CSC 201906100030). ZX would like to thank the helpful discussion with Honghui Liu.

\textit{Software}: XSPEC \citep{1996ASPC..101...17A}, Astropy \citep{2013A&A...558A..33A}, Numpy \citep{2011CSE....13b..22V}, Matplotlib \citep{2007CSE.....9...90H}, Stingray \citep{2019ApJ...881...39H,2019JOSS....4.1393H}.

\section*{Data Availability}

The raw \nicer\ data underlying this article are available in the HEASARC database.
The raw \hxmt\ data underlying this article are available at \url{http://hxmten.ihep.ac.cn/}.



\bibliographystyle{mnras}
\bibliography{mnras_template} 




\appendix
\section{Additional Figure}

\begin{figure*}
    \centering
    \includegraphics[width=0.98\textwidth]{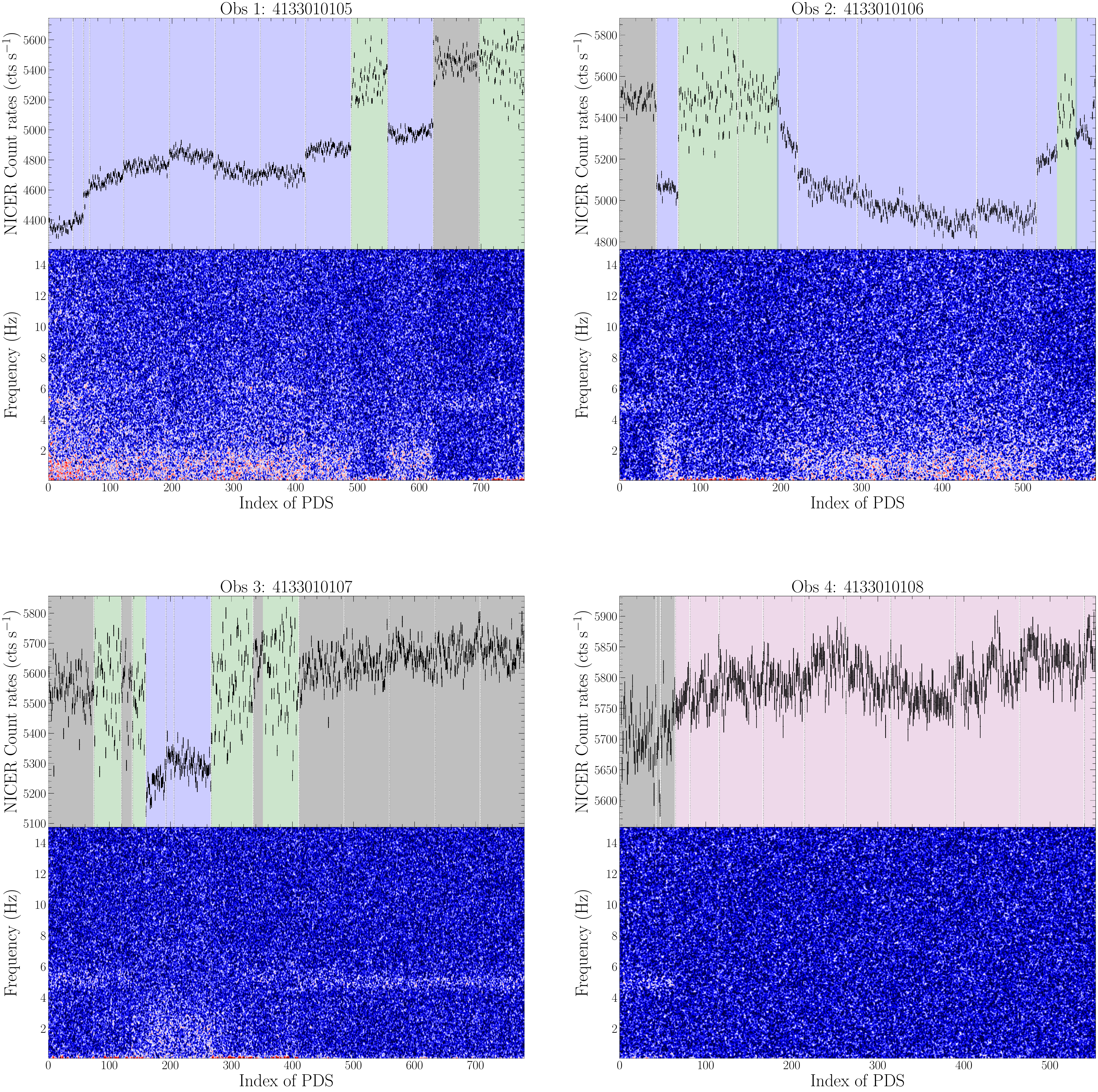}
    \caption{\nicer\ light curve (0.5--10 keV) (upper panels) and corresponding dynamical power spectra (lower panels) of \target\ with a time resolution of 16s. We divide the light curves into four typical periods based on the properties of PDS: periods with type-B QPO (gray shaded), with strong band-limited noise (blue shaded), with flip-flop phenomenon (green shaded) and the period dominated by Poisson noise (magenta shaded). }
    \label{lc_dps_appendix}
\end{figure*}


\bsp	
\label{lastpage}
\end{document}